# The Proton Beam Real-time Monitor System in CSNS

Jian ZHUANG[1,2,3], Jiajie LI[2,3], Ke ZHOU[2,3], Fang LI[2,3], Yongxiang QIU[2,3], Lei HU[2,3], ,

State Key Laboratory of Particle Detection and Electronics[1], Beijing, P.R.China
Institute of High Energy Physics[2], Beijing 100049, P.R.China
Dongguan Neutron Science Center[3], Dongguan 523803, P.R.China

*Abstract*–In Chinese Spallation Neutron Source (CSNS), proton beam is used to hit metal tungsten target, and then high flux neutron are generated for experiments on instruments. For neutron flux spectrum correction, the current of proton beam is for each instrument.

A real time monitor system is developed in CSNS, to monitor, broadcast and record each pulse of proton. Each proton pulse charge is measured and marked with high-precision timestamp. Then, the result of measurement will be broadcasted to control room and each neutron instrument. In control room, the proton charge of each pulse is listened and stored in database by agent program for offline use. The high-precision timestamp can be used to proton charge and neutron data alignment in time scale. The architecture of proton beam monitor system is introduced in this paper. And the performance of this system is evaluated in this paper.

## I. Introduction

CSNS mainly consists of an H-Linac and a proton rapid cycling synchrotron. It is designed to accelerate proton beam pulses to 1.6GeV kinetic energy at 25Hz repetition rate. Proton pulses hit a solid metal target to produce spallation neutrons [1]. The facility of CSNS is shown in figure Fig.1. There are several proton measurement device in RTBT, especially 3 beam current transformers.

The experimental control system of CSNS is in charge of target and instrument control [2][3]. The task of instrument control mainly includes providing local control to instrument device; integrating all devices belonged to the instrument into one system; providing facility information; providing the trigger signal (T0), and synchronizing all system.

In CSNS, the proton charge needs to be measured, to normalize neutron flux by the physical analysis software. The proton charge is also be used as option of termination conditions and statistical data of a neutron experiment. The time of proton bunch passing the beam monitor is also can be an option as the start of neutron flight time.

In this paper, the framework of proton beam real-time monitor system, including proton beam charge measurement, high-precision timestamp, real-time broadcast, listening agent, history database and statistics in introduced. The running result of this proton real-time monitor system in CSNS instrument commissioning is also introduced in this paper.

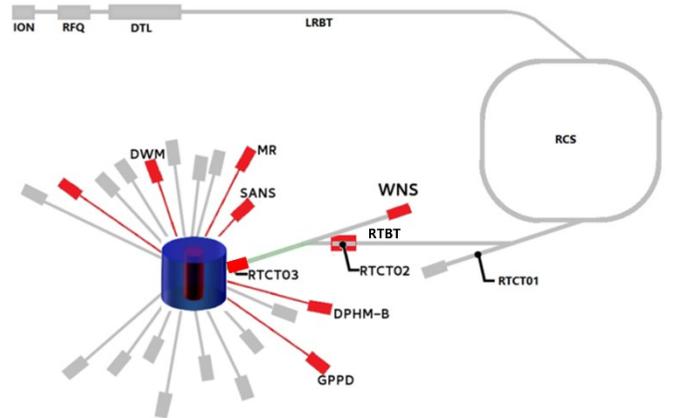

Fig.1. Overall View of CSNS

## II. Framework of Proton Beam Real-time Monitor System

The proton beam is measured by the proton current transformer in RTBT before the target. A time synchronization and real-time data network based on White Rabbit is estabilshed in CSNS. When a proton bunch passing, the TDC electronics receive a trigger signal from the probe of proton beam monitor and mark it with high-precision timestamp. Then, TDC card will send a trigger message with this timestamp through optical fiber Ethernet for high time restriction application. Then the proton beam Charge is measured by proton current transformer. Because of the latency of ADC, the measurement process will be completed later. The proton charge is marked with the same timestamp as trigger and the measurement message is broadcasted through optical fiber Ethernet.

To utilize these message, there are specific agents listening the broadcast message. Two redundant agents are deployed to save each proton charge into MySQL database. To ensure that the proton query operation do not affect writing the proton message to database, the read and write separation technology is adopted. And a web page is developed to help query these proton message easily.

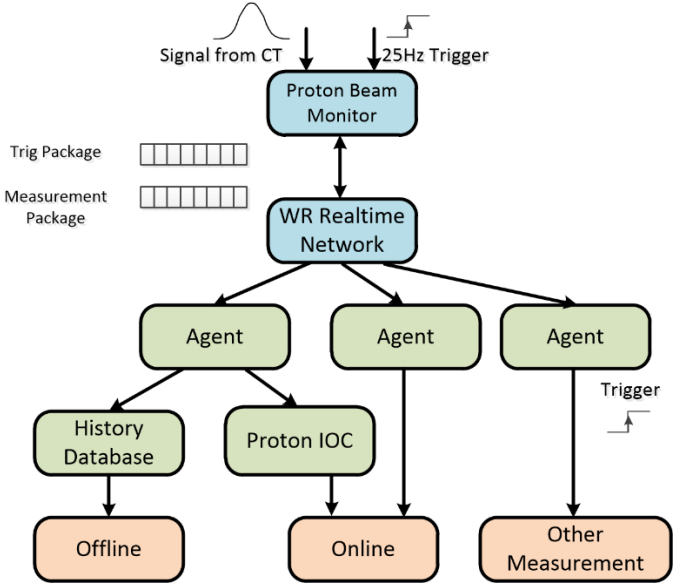

Fig.2. Framework of NEROS

There are two ways to use proton message for online. A proton message IOC with agent is used to provide PVs of proton charge and power to online software. In order to ensure that there are no large amounts of data traffic injected into real-time network to affect the performance of real-time network, agents are used to isolate the data communication network and real-time network. For physical online analysis, a special agent is developed to provide the proton message in last minute.

The timestamp precision of proton is about 1ns, and it is enough to neutron instrument for the reason that cold and thermal neutrons are mainly used in instruments of CSNS. The real-time message can be faster than cold and thermal neutrons in almost all neutron instrument. A dedicate hardware agent is developed to listening the real-time trigger message, and fan out a trigger signal in determined time point later. This can simplify the design of T0 and help measure the TOF of neutron more accuracy in the future.

### III. TIME SYNCHRONIZATION

To mark high precision timestamp on proton beam message, a high time precision synchronization network based White Rabbit Technology is established in CSNS [4][5][6], as shown in Fig. 3. White Rabbit Technology is originated from CERN, and uses a specific protocol improved from IEEE 1588 and the optical synchronous Ethernet to realize time synchronization system with high precision.

A GPS receiver, rubidium clock and WR grandmaster switch are deployed in the control room. The WR grandmaster is synchronized to GPS by pulse per second (PPS). A rubidium is used to acclimate PPS signal from GPS receiver and providing stable 10MHz clock. The UTC time is also get from GPS receiver by NTP service.

Several devices based on WR technology including PXI TDC card, PXI delay card, timing fanout box and so on are developed for measurement and integration of different electronics. For the electronics not based on WR technology, a timing fanout chassis is developed to generate a PPS signal with a series pulse standing for TAI time. So different electronics with different crystal oscillator can track this PPS signal to be synchronized. The timestamp over second is come from WR synchronization. The timestamp under second is come from crystal oscillator of electronics itself, and it will be formatted into nanosecond. Each event in the neutron experiment can be marked high precision timestamp and serialized by time sequence.

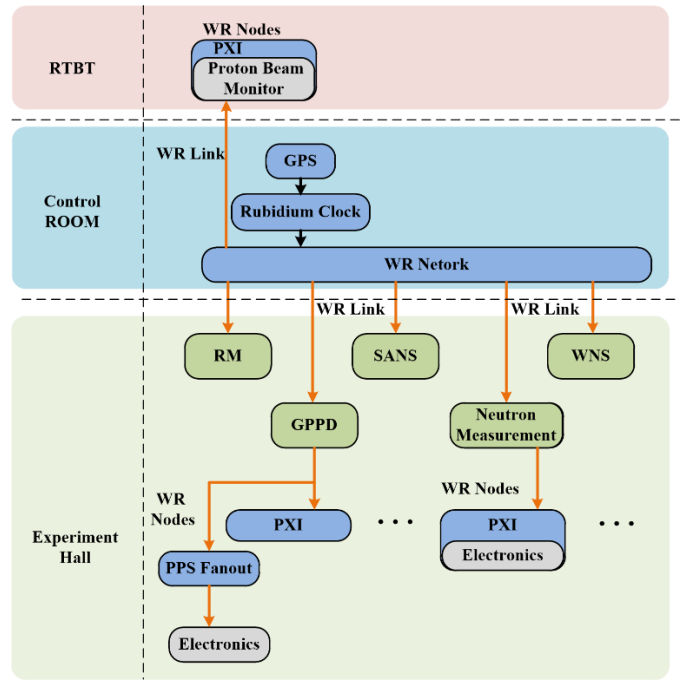

Fig. 3 Time Synchronization

### IV. PROTON BEAM MEASUREMENT

There are 3 proton CT (current transforms) probe in RTBT section of CNS. RTCT02 and RTCT03 are used to monitor the proton beam. The electronics of proton beam monitor is based on NI PXI chassis with PXI controller installed. There are NI PXI oscilloscope card and self-developed TDC card in the chassis. Beam current transformer and beam position monitor are used to monitor proton beam.

When a proton bunch passing, the induced current is generated on the primary coil of CT probe. The voltage on resistor connected to secondary coil can be sampled by ADC card. The signal sampled is filtered through the Butterworth filter in order to remove noise. The background is removed by the baseline deduction. The Savitzky-Golay filter is used to smooth the curve. Finally the charge of proton bunch is calculated by integration of the sampled signal. Integration window is adjusted carefully to ensure the best SNR. All the data processing is done in the PXI controller.

## V. REAL-TIME BROADCAST MESSAGE

A real-time message derived from LXI protocol is defined in CSNS for real time measurement, as shown in table 1. Fields such as sequence and timestamp are filled by FPGA. The other fields are set by software in configuration phase.

Table 1 Realtime Trigger Message Type

| Name | Field | Type | Length |
|---|---|---|---|
| Domain | HW_Detect | string | 3 Bytes |
|  | Domain | unit8 | 1 Byte |
| Event ID | Trigger Type and signal ID, hardware and Channel ID | string | 16 Bytes |
| Serial number | Sequence No. | unit32 | 4 Bytes |
| Timestamp (Trigger time) | seconds | uint32 | 4 Bytes |
|  | nanoseconds | uint32 | 4 Bytes |
|  | fractional_nanoseconds | uint16 | 2 Bytes |
|  | Epoch | uint16 | 2 Bytes |
| Flags | Flags | uint16 | 2 Bytes |
| Data Field | Data Field |  | 66 Bytes |
| END | End (0x00) | uint16 | 2 Bytes |

The domain field is set by measurement software, typical as EXP to identify neutron experiment. Event ID is also set by the software. Event ID is divided in to several sub-fields and a coding rule is formulated to identify measure type, signal ID, device ID and Channel ID. For the proton beam monitor in CSNS now, there are two types of real-time message to deliver trigger and measurement result. And the "proton_TRG" is for trigger message and "proton_Data" for measurement message. The field of serial number is set by FPGA on the PXI card and will be plus 1 automatically for each trigger. The timestamp field records the time point was trigged. This field is divided into 4 parts, including second, nanosecond, fractional of nanoseconds and epoch. TAI time is used in second field. According to LXI and IEEE 1588 protocol, the most significant 16 bits of timestamp is set in epoch field. Flag field is reserved for compatibility with the LXI protocol. For different message, the data field is different. At last, the message is end with 0x00.

Considering the long-term running of CSNS, 64 bits width is more suit for the serial number and timestamp. So, for the trigger message, there are optional serial number and trigger timestamp in data field reserved. The serial number and trigger timestamp of measurement are set as same as trigger message in one measurement. Compared with trigger message, there are 3 proton charge field more in measurement message. Each data in the data field has 3 sub-fields. Data Length indicates the length of data, and identifier indicates the type of data. The typical value of these two field is compatibility with LXI protocol standard.

Table 2 Detail in Data Field

| Field | Type | Length | Typical Vaule |
|---|---|---|---|
| Data Length | uint16 | 2 Bytes | 0x04 |
| Identifier | unit8 | 1 Bytes | 0xf7 |
| Proton bunch Seqence | unit32 | 4 Bytes |  |
| Data Length | uint16 | 2 Bytes | 0x08 |
| Identifier | unit8 | 1 Bytes | 0xf9 |
| Proton Trigger Time Second | unit64 | 8 Bytes |  |
| Data Length | uint16 | 2 Bytes | 0x04 |
| Identifier | unit8 | 1 Bytes | 0xf7 |
| Proton Trigger Time Nanosecond | unit32 | 4 Bytes |  |
| Data Length | uint16 | 2 Bytes | 0x04 |
| Identifier | unit8 | 1 Bytes | 0xf6 |
| Proton Charge ct01 | unit32 | 4 Bytes |  |
| Data Length | uint16 | 2 Bytes | 0x04 |
| Identifier | unit8 | 1 Bytes | 0xf6 |
| Proton Charge ct02 | unit32 | 4 Bytes |  |
| Data Length | uint16 | 2 Bytes | 0x04 |
| Identifier | unit8 | 1 Bytes | 0xf6 |
| Proton Charge ct03 | unit32 | 4 Bytes |  |

## VI. AGENT

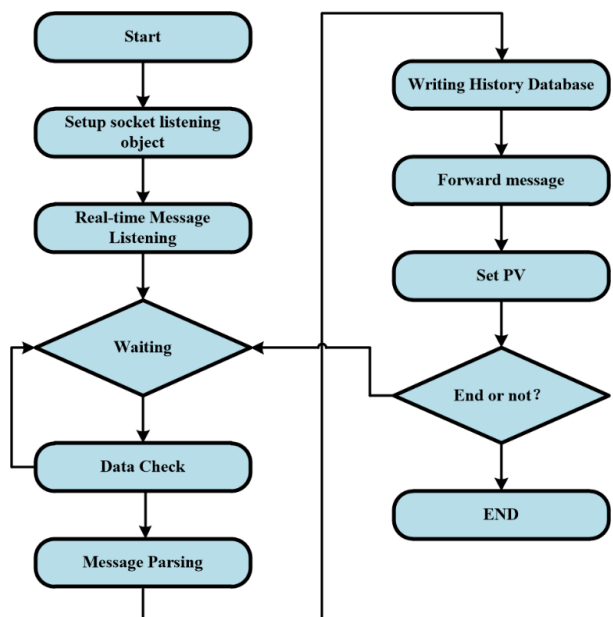

Fig. 4 Software flow of Agent sofware

To isolate the WR real-time network from control and experiment data network, agent programs based on Python and C language are developed and double network card servers are used. One network card is connected to WR real-time network, and the other is connected to experiment data network. The agent program is running as a service of Linux, listening the message continuously from proton beam monitor, as shown in Fig. 4.

After the message received, the data format is check first. Then, the message is parsed to get the proton charge, serial number and timestamp. Agent program will forward these value to experiment data network, and store these value to history database.

## VII. OFF-LINE DATA CHECK AND PUBLISH

To check the validation and reliability of proton beam data, a Python program is developed to check the data stored in database periodically, and publish outside.

First, the charge of proton beam will be checked, and the data lower than $0.2 \times 10^{12}$ will be filtered. Then, the sequence and time interval will be check and the abnormal records will be printed for hint.

## VIII. RUNNING ON CSNS INSTRUMENT

From Nov.1 2017 to Feb. 28 2018, CSNS has completed the joint commissioning of accelerator, target and neutron instruments. In this period, 5.4MWhr proton power was accumulated on the target, as shown in Fig. 5. There are $3.9 \times 10^7$ pulses of proton accumulated on target in these 4 months.

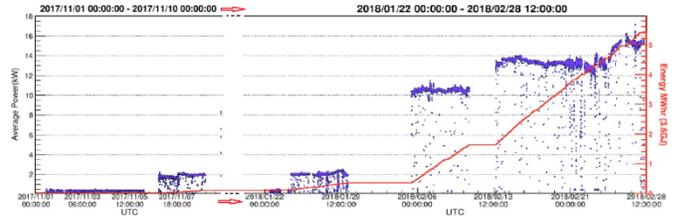

Fig. 5 Proton Power on Target

## IX. CONCLUSION

In the Chinese Spallation Neutron Source, a proton beam monitor based on WR time synchronization is developed. A TDC card and real-time message based on LXI protocol are also developed and deployed. Several agent program based on Python different online and offline use. Through 4 months commissioning in CSNS, the system can achieve the goal of design, and there are no data missing found in $1 \times 10^6$ pulsed.

## X. ACKNOWLEDGEMENTS

This work was supported by China Spallation Neutron Source. This work is also was support by the science and technology project of Guangdong province under grand No. 2016B090918131 and 2017B090901007.

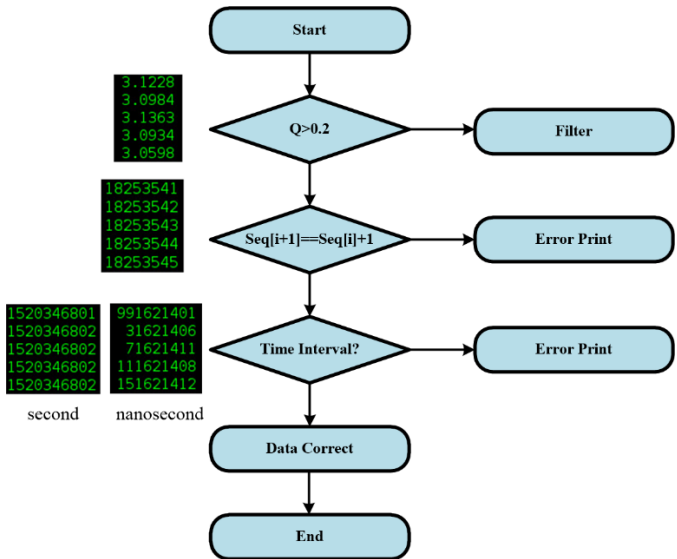

Fig.5 Check Flow of Proton Beam Data